\documentclass[aip, pop,  reprint]{revtex4-1}

\usepackage{amssymb}
\usepackage{amsmath}
\usepackage{graphicx}

\newcommand{\pderv}[2]{\frac{\partial #1}{\partial #2}}

\newcommand{\derv}[2]{\frac{{\rm d} #1}{{\rm d} #2}}
\newcommand{\dervi}[2]{{\rm d} #1/{\rm d} #2}
\newcommand{\rem}[1]{}
\def\rmi{\mathrm{i}}

\begin{document}
\title{Kinetic instabilities in a mirror-confined plasma sustained by high-power microwave radiation}
\author{A. G. Shalashov}
\email[Author to whom correspondence should be addressed. Electronic mail:\;]{ags@appl.sci-nnov.ru}
\author{M. E. Viktorov}
\author{D. A. Mansfeld}
\author{S. V. Golubev} 
\affiliation{Institute of Applied Physics of the Russian Academy of Sciences, 46 Ulyanov str., 603950 Nizhny Novgorod, Russia}

\date{\today}

\begin{abstract}
This paper summarizes the studies of plasma kinetic instabilities in the electron cyclotron frequency range carried out over the last decade at the Institute of Applied Physics in Nizhny Novgorod. We investigate the nonequilibrium plasma created and sustained by high-power microwave radiation of a gyrotron under the electron cyclotron resonance condition. Resonant plasma heating results in the formation of at least two electron components, one of which, more dense and cold, determines the dispersion properties of the high-frequency waves, and the second, a small group of energetic electrons with a highly anisotropic velocity distribution, is responsible for the excitation of unstable waves. 
Dynamic spectra and the intensity of stimulated electromagnetic emission are studied with high temporal resolution. 
Interpretation of observed data is based on the cyclotron maser paradigm, in this context, a laboratory modeling of non-stationary wave-particle interaction processes have much in common with similar processes occurring in the magnetosphere of the Earth, planets, and in solar coronal loops.
\end{abstract}


\maketitle

\section{Introduction}

Electron cyclotron instabilities of magnetically confined nonequilibrium plasma are common in space plasma. Most vivid manifestations are electromagnetic emissions in magnetospheres of the Earth\cite{trakh_books,chorus}, other planets\cite{Menietti_2012_Jupiter_Saturn} and stars \cite{Trigilio_2011_stars}. Precipitations of energetic particles, frequently accompanying such instabilities, affect the dynamics of the Earth's radiation belts \cite{Thorne_2010_RB}. Research of similar processes in laboratory conditions is very important for understanding the physical mechanisms of instabilities in space plasmas\cite{viktorov_EPL,Bingham_2013,VanCompernolle_2016}. 
Besides that, cyclotron instability is an important channel for the loss of excess energy stored in the fast particles \cite{Garner_1990,viktorov_rf1}, thereby limiting the achievement of peak plasma parameters in applications \cite{olli_2015}.

In this paper we review the kinetic instabilities of nonequilibrium plasma supported by high-power microwave radiation under the electron cyclotron resonance (ECR) condition in an axisymmetric open magnetic trap. Instabilities are manifested as a generation of short pulses of electromagnetic emission in electron cyclotron (EC) frequency range accompanied by precipitation of hot electrons from the trap. We study detailed time-frequency characteristics of the stimulated electromagnetic emission, what has been made possible only recently with the advent of methods for measuring the electromagnetic field with high temporal resolution. With respect to the other laboratory studies of EC instabilities \cite{Bingham_2013,VanCompernolle_2016}, where plasma emissions are excited by electron beams, we explore microwave emissions generated by much wider distributions of fast electrons in a velocity space driven by strong ECR heating of target plasma.

\section{Experimental setup and conditions}

\begin{table*}[tb]
\caption{ \label{tab:plasma_parameters}Plasma parameters at different stages of ECR discharge.}
\begin{ruledtabular}
\begin{tabular}{@{}lllll}
 & \textbf{ECR start-up}  & \textbf{Stationary discharge}  & \textbf{ECRH switch-off}  & \textbf{Plasma decay}   \\
 & $N_{\mathrm{h}} \gtrsim N_{\mathrm{c}}$ & $N_{\mathrm{h}} \ll N_{\mathrm{c}}$ &$N_{\mathrm{h}}\lesssim N_{\mathrm{c}}$& $N_{\mathrm{h}}\sim N_{\mathrm{c}}$\\
 \hline
 Cold electron density $N_{\mathrm{c}}$&  --  &  $\sim 10^{13}$ cm$^{-3}$  &$\sim 10^{12}$ cm$^{-3}$ & $\lesssim 10^{11}$ cm$^{-3}$\\
 Cold plasma temperature $T_c$ &  --  & $100-300$ eV &$\sim 100$ eV& $\sim 1$ eV \\ 
 Hot electron density $N_{\mathrm{h}}$ & $\sim 10^{10}$ cm$^{-3}$ & $10^{11}$ cm$^{-3}\;/\;\; 10^{9}-10^{10}$ cm$^{-3}$&$\sim10^{11}/\,10^{10}$ cm$^{-3}$& $\sim10^{11}/\,10^{10}$ cm$^{-3}$\\
 Hot electron energy $\varepsilon_{\mathrm{h}}$ & $\sim 300$ keV & $\sim 10$ keV $\;\;/\;\; 10-100$ keV & $ \sim10/\,100$ keV & $\sim10/\,100$ keV \\ 
 Plasma frequency to minimum   \\ 
 cyclotron frequency   $\omega_{\mathrm{pe}}/\omega_{\mathrm{ce}}$ & $\omega_{\mathrm{pe}}/\omega_{\mathrm{ce}}\ll 1$ & $\omega_{\mathrm{pe}}/\omega_{\mathrm{ce}}\gg 1$ &$\omega_{\mathrm{pe}}/\omega_{\mathrm{ce}}\gtrsim 1$& $\omega_{\mathrm{pe}}/\omega_{\mathrm{ce}}\ll 1$ \\
\end{tabular}
\end{ruledtabular}\end{table*}

\begin{figure}[b]
\centering
\includegraphics[width=83mm]{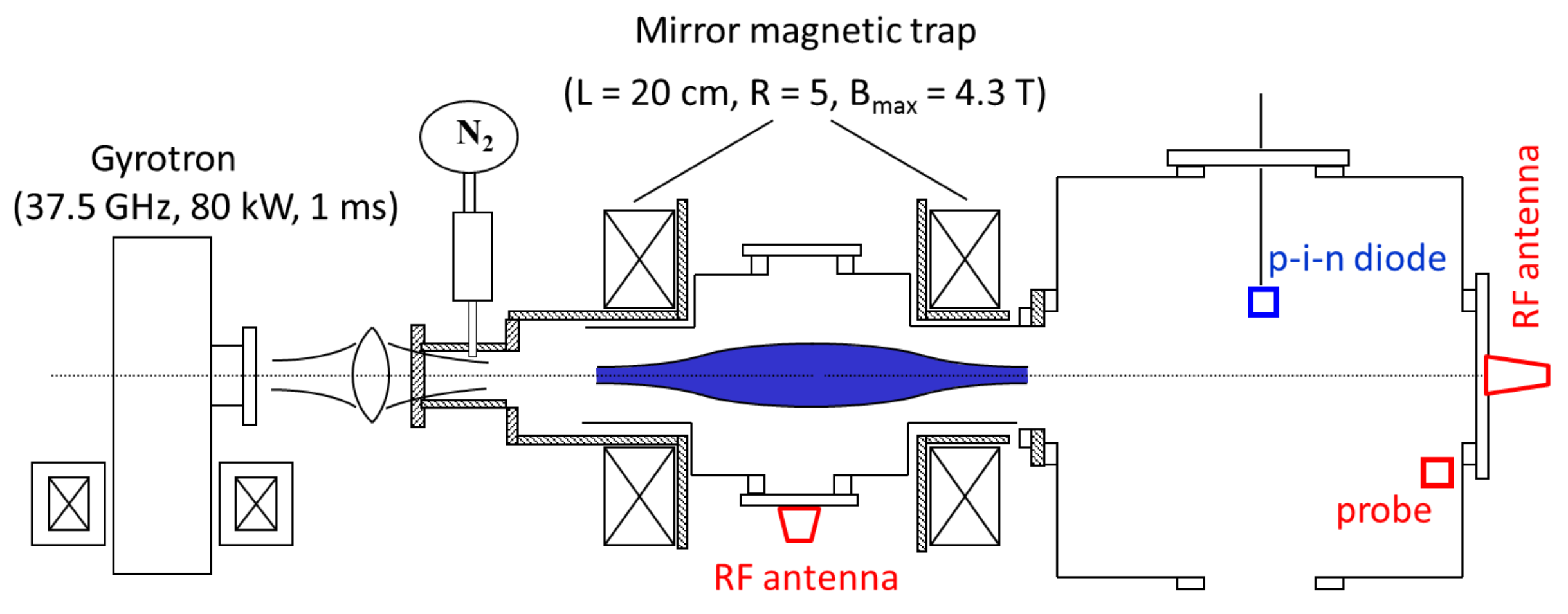}
\caption{The schematic view of the SMIS-37 setup.}
\label{fig:setup}
\end{figure}

The experiments were performed in the plasma of ECR discharge sustained by high-power millimeter-wave gyrotron radiation (at frequency 37.5 GHz, power up to 80 kW, pulse duration up to 1 ms)  in the compact magnetic mirror trap SMIS-37 \cite{Mansfeld_2007_JETP_maser}. The schematic view of setup is shown in Fig.~\ref{fig:setup}. 
The axially symmetric discharge chamber is placed in the mirror magnetic trap of the length 200\,mm, produced by pulsed coils which allow obtaining maximum magnetic field strength of 4.3 T, mirror ratio is about 5, pulse duration 7\,ms. The discharge chamber is a tube with inner diameter of 38\,mm which is widened in center part by the tube with inner diameter of 72\,mm and 50\,mm length. Plasma is created and supported under the ECR conditions at the fundamental cyclotron harmonic. The microwave radiation is launched along the trap axis through the teflon window and is focused to one of the ECR zones by a matching unit. The ECR surfaces, which correspond to the magnetic field strength 1.34\,T, are situated symmetrically between the magnetic mirror point and the trap center. Ambient pressure of a neutral gas varies during experimental shot from $10^{-6}$ to $10^{-3}$\,Torr. Most of the reported experiments are performed in nitrogen, although similar results are obtained in argon and other gases.

At least four stages of ECR discharge with different plasma parameters can be identified \cite{viktorov_EPL}, see table~\ref{tab:plasma_parameters}. At the first stage, with a duration of about 100 $\mu$s, the plasma density is small, and the absorbed microwave energy is enough to accelerate electrons up to relativistic energies. At the second stage, which lasts until the end of the microwave pulse, the bulk plasma density is higher than during the first stage by more than two orders of magnitude. At that period, the plasma consists of two-components, a cold dense component ($N_{\mathrm{c}}$, $T_c$) with an isotropic velocity distribution, and the less dense component of hot electrons ($N_{\mathrm{h}}$, $\varepsilon_{\mathrm{h}}$) with anisotropic distribution function\cite{avod1999,avod_whist}.
The transition from the initial start-up discharge stage to the stage of developed ECR discharge takes about several decades of microseconds and is related to the plasma density increase resulted from ionization of neutral gas arriving into the trap with increasing flow rate due to desorption from the inner walls \cite{Vodopyanov_RQE_2003}. 
The energy stored in hot electrons is partially spent on ionization of the incoming neutral gas. During this stage, the plasma density is about the critical density $1.7\times 10^{13}$\,cm$^{-3}$ for microwave radiation at 37.5\,GHz. At this density level,  the absorbed microwave power  is not sufficient to heat an essential number of electrons to relativistic energies; thus the average electron energy decreases.
The third stage starts right after the ECR heating switch-off, when the density of cold fraction decreases rapidly, while the hot electrons with an anisotropic velocity distribution function are confined in the magnetic trap much longer. Starting from a certain time, the density of the hot component become equal to or even higher than the density of the cold component. The fourth stage is observed with a delay about hundreds of microseconds in a decaying plasma when the density of cold plasma is much less than the density of the hot component.

The dynamic spectrum and intensity of stimulated plasma emissions are studied with use of high-performance oscilloscopes (Tektronix MSO\,72004C with analog bandwidth 20\,GHz,  sampling rate 100\,GSample/s, and Keysight DSA-Z\,594\,A with analog bandwidth 59\,GHz, sampling rate 160\,GSample/s). 
The signal is picked up with a broadband horn antenna with a uniform bandwidth in the range from 2 to 20\,GHz. The directional antenna is placed outside the discharge chamber near one of two vacuum windows as shown in Fig.~\ref{fig:setup}. Thus, in principle,  we can detect the microwave emission propagating along and across the magnetic field separately. 
However, in the reported experiments the level of signal in both directions was essentially similar in most regimes.
Most reliable data on the propagation direction   was obtained in earlier experiments performed with more robust equipment.

To cover all stages of ECR discharge, the length of the recorded waveforms is set to 5\,ms what corresponds to $(5-8)\times 10^8$ data points (depending on oscilloscope model) per channel for a single experimental shot.
The dynamic spectra are calculated from the recorded oscilloscope data by short-time Fourier transform with a Hamming window. The window size $\tau_{\mathrm{w}}$ is defined to fulfill the following conditions:
$$
\tau_{\mathrm{sw}}^{-1}<\tau_{\mathrm{w}}^{-1}\ll f \lesssim f_{\mathrm{s}}/2,
$$
where $\tau_{\mathrm{sw}}$ is the characteristic time of frequency sweeping, $f$ is the carrier frequency of the received signal (of the same order as the  cyclotron frequency $f_{\mathrm{ce}}=\omega_{\mathrm{ce}}/2\pi\sim6-10$\,GHz or its lower harmonics), and $f_{\mathrm{s}}$ is the oscilloscope sampling rate. 

Simultaneously to the plasma emission in  $2-20$ GHz frequency band, we measure precipitations of energetic electrons ($>10$\,keV) from the trap ends by a pin-diode detector with time resolution about 1\,ns. 

This experimental technique was previously reported in Refs.~\onlinecite{viktorov_rf2, viktorov_rf_dpr, viktorov_EPL}. As compared to our early studies
, it allows us to detect the instabilities during all stages of a discharge in one experimental shot and distinguish fine details of the radiation spectrum. 

\section{Microwave emissions of mirror-confined plasma}

\begin{figure*}[t]
\includegraphics[width=0.8 \textwidth]{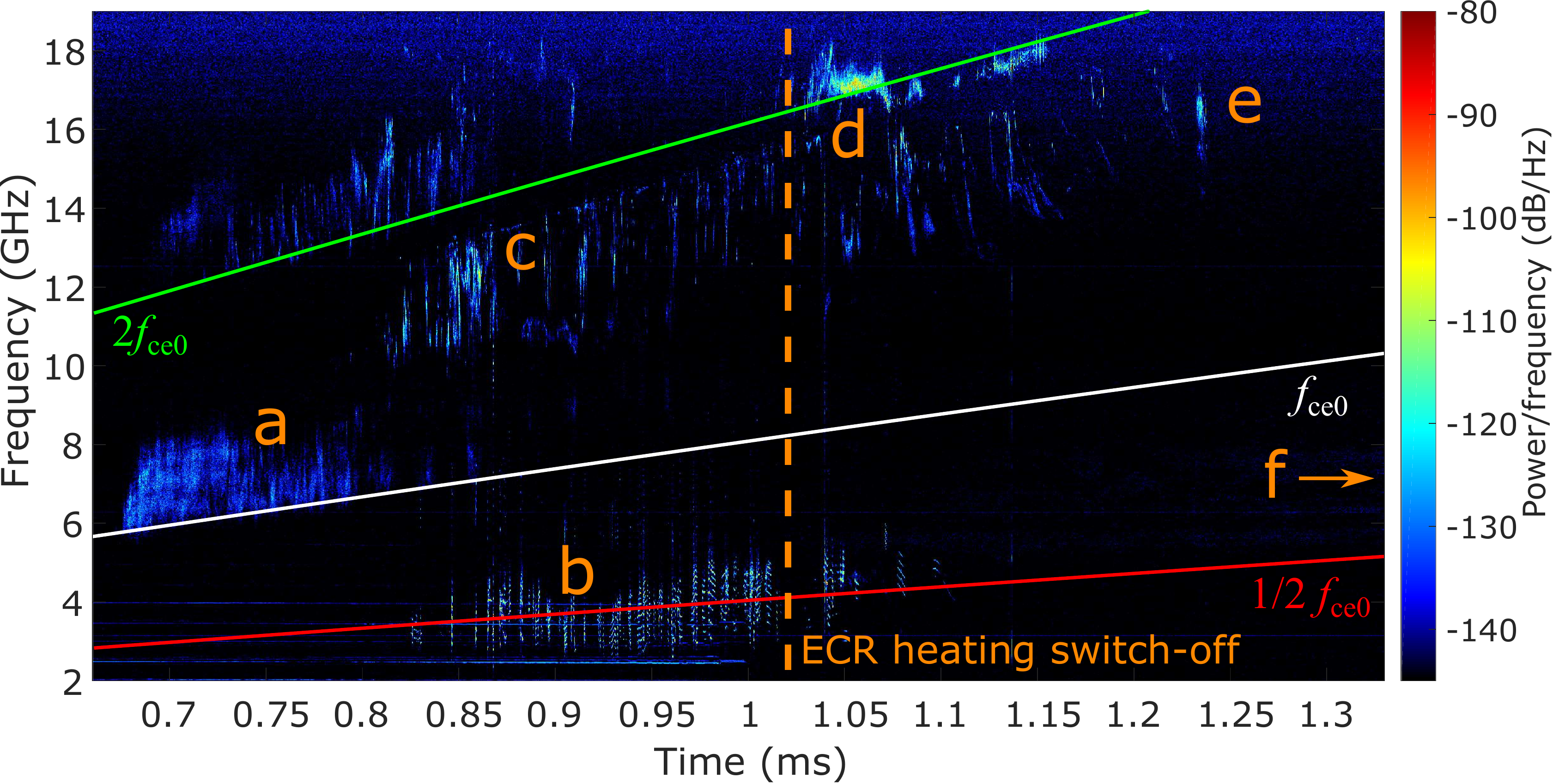}
\caption{Example of dynamic spectrum of the plasma emission during and after ECR discharge. Instabilities of type (a)-(e) are indicated, type (f) is out of the time slot. Straight lines show the variation of $2f_{\mathrm{ce0}}$ (green), $f_{\mathrm{ce0}}$ (white) and $f_{ce0}/2$ (red), where $f_{\mathrm{ce0}}$ is the electron cyclotron frequency in the center of the magnetic trap on its axis. The ECR heating is switched-on at 0 ms (not shown) and switched-off at 1 ms. The left boundary of the time axis corresponds to ECR breakdown when the plasma becomes visible in all diagnostics. The signal is detected by $2-20$\,GHz antenna  looking along to the magnetic field.}
\label{fig_spectrum}
\end{figure*}

The described stages of pulsed ECR discharge offer the opportunity to simultaneously study wave-particles interactions for essentially different plasma parameters: at the initial stage, when the density of hot relativistic electrons exceeds the density of cold electrons, at the developed steady ECR discharge, and in the decaying plasma after the gyrotron switch-off. On each stage, we detect  series of quasi-periodic broadband pulses of electromagnetic radiation with a typical duration of a few microseconds, and related precipitations of energetic electrons, which are presumably caused by cyclotron instability of different electromagnetic modes or by generation of plasma waves. It should be stressed here that kinetic instabilities provide the only reasonable mechanism for losses of adiabatically confined collisionless hot electrons in our experiment.

A typical dynamic spectrum of the electric field oscillations in the excited wave is shown in Fig.~\ref{fig_spectrum}. This example gives a good overview since  almost all types of instabilities may be observed simultaneously.
Different  kinds of  instabilities are considered below in more detail. Note, that particular details of the measured electromagnetic spectrum may vary from one shot to another while all other controlled parameters of the setup remain unchanged. 
Most of observed dynamic spectra are strongly correlated to the time-varying electron cyclotron frequency $f_{\mathrm{ce0}}$ at the trap axis in the central cross-section. That evidences in favor that the observed radiation is excited under electron cyclotron conditions and the region of effective wave-particle interaction is more likely situated in the trap center, where the magnetic field is the most uniform. Location of the  effective interaction area in the region of the most homogeneous magnetic field was discussed earlier for space cyclotron masers.

\subsection*{(a)  X-mode emission at the start-up phase }

\begin{figure*}[tp]
	\includegraphics[width=1.0 \textwidth]{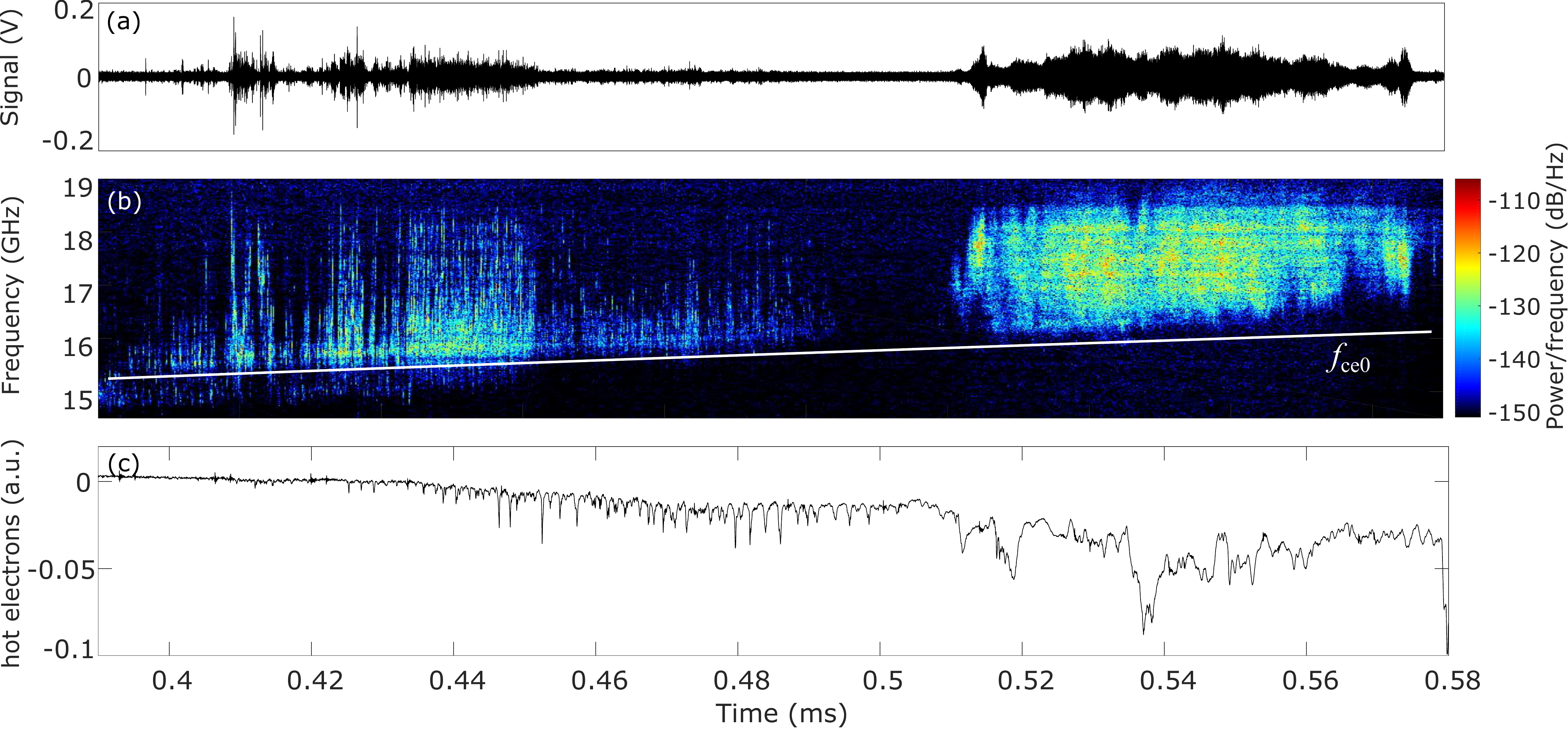}
	\caption{(a) Waveform and (b) its dynamic spectrum of plasma microwave emission at the start-up  stage,  (c) the corresponding signal from the hot electron detector. The value of the electron gyrofrequency $f_{\mathrm{ce0}}$ at the trap center is shown by a white solid curve on the dynamic spectra.}
	\label{fig_spectrum1}
\end{figure*}
\begin{figure*}[tph]
	\includegraphics[width=0.95 \textwidth]{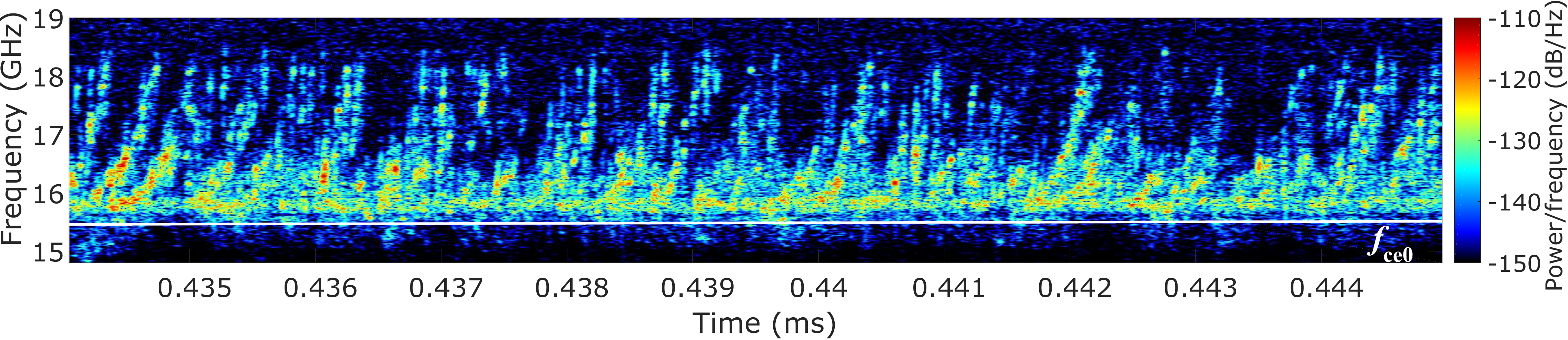}
	\caption{Zoom of spectrum shown in Fig.~\ref{fig_spectrum1}(b). Here the periodic fast rising tones are clearly visible. }
	\label{fig_spectrum1_zoom}
\end{figure*}

At the start-up phase of the ECR discharge while dense plasma is absent  we detected microwave radiation in a direction presumably perpendicular to the ambient magnetic field at frequencies slightly higher than the electron cyclotron frequency $f_{\mathrm{ce0}}$, see Fig.~\ref{fig_spectrum} label \textit{a}. Therefore, it is more naturally explained by the excitation of the loss-cone instability of a fast extraordinary (X) wave, which  propagates obliquely to the magnetic field \cite{shalash_prl}. 

We explored two significantly different regimes of instabilities -- periodic wave packets with fast raising tones smoothly transforms into emission with the continuous spectrum \cite{viktorov_rf2}. This transition is shown in Fig.~\ref{fig_spectrum1}. During time 400-500\,$\mu$s periodic emission of fast rising tones is observed; this is visible in Fig.~\ref{fig_spectrum1_zoom} which shows a zoom of the same spectrum as Fig.~\ref{fig_spectrum1}. 
The frequency sweeping rate in a single wave packet is about 5\,GHz/$\mu$s. Then the registered signal is almost zero for a period of about 10\,$\mu$s. After that, the spectrum of microwave emission is changed to the continuous broadband radiation during 510-580\,$\mu$s. In both cases, the bandwidth of emission is about several GHz. It is seen from the spectrogram that the lower spectrum boundary of radiation is greater than $f_{\mathrm{ce0}}$ by an amount of about $0.01f_{\mathrm{ce0}}$. 

Analysis of the most favorable conditions for excitation of the loss-cone instability may be used to estimate the maximum energy of the hot resonant electrons \cite{viktorov_rf2}. This energy is related to the upper boundary of the observed emission frequency as  $\varepsilon_{\mathrm{h}}\approx(\omega_{\max}/\omega_{\mathrm{ce}}-1)\,m_{\mathrm{e}} c^2$. For the case shown in Fig.~\ref{fig_spectrum1}, $\omega_{\max}/\omega_{\mathrm{ce}}\approx 1.2$, which indicates that electrons are accelerated up to 100\,keV at the start-up discharge stage. In other experiments the ratio $\omega_{\max}/\omega_{\mathrm{ce}}$ may be up to 1.6, which corresponds to the energies up to 300\,keV. This agrees well with the previously published data on the measurement of the electron energy distribution \cite{Izotov_2012_RSI}.

\subsection*{(b) Whistler waves during the stationary ECR discharge}

\begin{figure*}[tp]
	\includegraphics[width=1.0 \textwidth]{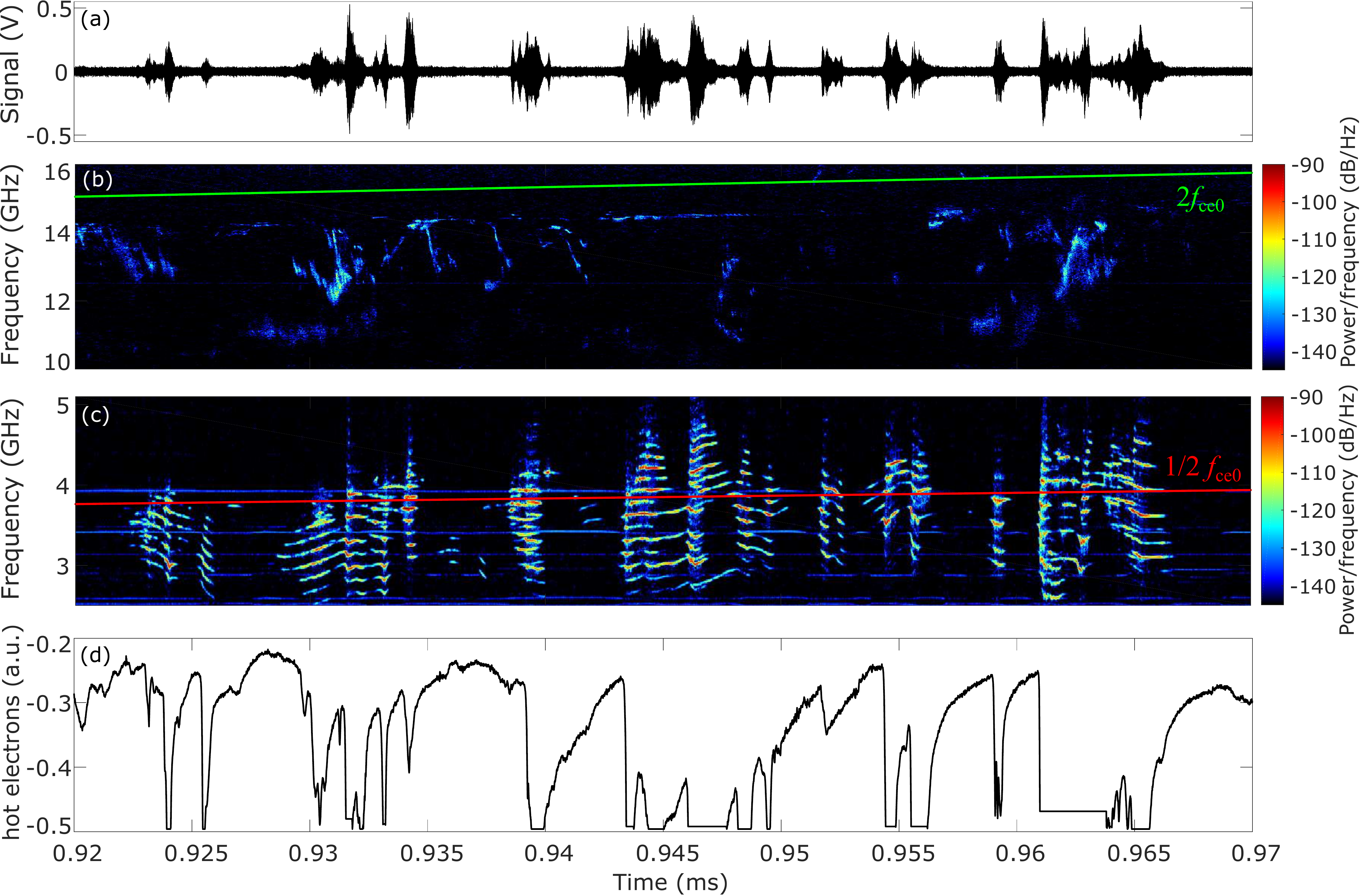}
	\caption{(a) Waveform and (b,c) its dynamic spectrum of plasma microwave emission during the stationary ECR discharge, (d) the signal from the hot electron detector. The values of $f_{\mathrm{ce0}}/2$ and $2f_{\mathrm{ce0}}$ at the trap center are shown by solid curves on the dynamic spectra.}
	\label{fig_spectrum2_3}
\end{figure*}

During the developed discharge phase we registered microwave emission in a direction along the trap axis, i.e. longitudinal to the ambient magnetic field, at frequencies about $f_{\mathrm{ce0}}/2$, see Fig.~\ref{fig_spectrum} label \textit{b}. Figure \ref{fig_spectrum2_3} shows an example of such quasi-periodic bursts of radiation in more detail. 
Every radiation pulse is strongly correlated with precipitations of energetic electrons  measured by the p-i-n detectors. The duration of pulses is about 1\,$\mu$s. The distinctive feature of this type of instability is the presence of the selected frequencies (more than ten) in the spectrum, which are arranged equidistantly relatively to each other. These frequencies of spectral components are slightly changing in time while the distance between them is constant. Typical distance between spectral components is $\Delta f \approx 150$ MHz ($\Delta f/f\approx 0.05$) and the spectral width of a single component is $\Delta f_{\mathrm{single}} \approx 20$ MHz ($\Delta f_{\mathrm{single}}/f\approx 7\times 10^{-3}$), see Fig.~\ref{fig_spectrum2_3}c.

At a large density of the background plasma during the stationary ECR discharge  stage, $\omega_{\mathrm{pe}}\gg\omega_{\mathrm{ce}}$, cyclotron instabilities of the extraordinary waves are suppressed, because their dispersive properties are strongly modified by the background plasma. Emission of dense plasma at frequencies about $f_{\mathrm{ce0}}/2$ is most naturally related to the whistler mode instability. At our setup, this instability was first  observed and explained  based on rather crude data obtained with microwave detectors and a filter bank \cite{avod_whist}. Later,  essentially more accurate dynamic spectra were derived using the abilities of a wideband digital oscilloscope \cite{viktorov_EPL}. New detailed experimental data seems to be in full agreement with the original interpretation of the observed instability below $f_{\mathrm{ce0}}$ as excitation of whistler modes trapped inside the plasma column.

\subsection*{(c) Microwave emissions above $f_{\mathrm{ce}}$ in dense plasma}

Simultaneously to the whistler wave instability, the other instability was observed at frequencies between $f_{\mathrm{ce0}}$ and $2f_{\mathrm{ce0}}$, see Fig.~\ref{fig_spectrum} label \textit{c}. 
The propagation direction of this radiation can not be clearly identified. 
The dynamic spectrum has a sharp upper boundary which varies in proportion to the frequency of $2f_{\mathrm{ce0}}$. This microwave emissions apparently can be related to the excitation of plasma waves under the upper hybrid resonance (UHR). Plasma waves propagate perpendicular to the ambient magnetic field and then transform to electromagnetic waves (e.g. slow extraordinary waves) in rarefied plasma at the boundary of the plasma column.
Figure \ref{fig_spectrum2_3}b shows a typical example of the high-frequency emission coming from the dense plasma. The upper cut-off frequency $f_{\ast}$ is clearly observed and $(2f_{\mathrm{ce0}}-f_{\ast}) \approx 0.1f_{\mathrm{ce0}}$. The spectrum of radiation is rather complex, but we can distinguish single wave packets with decreasing frequency. 

\subsection*{(d) Excitation of plasma waves under the double plasma resonance condition }

\begin{figure*}[tp]
	\includegraphics[width=1.0 \textwidth]{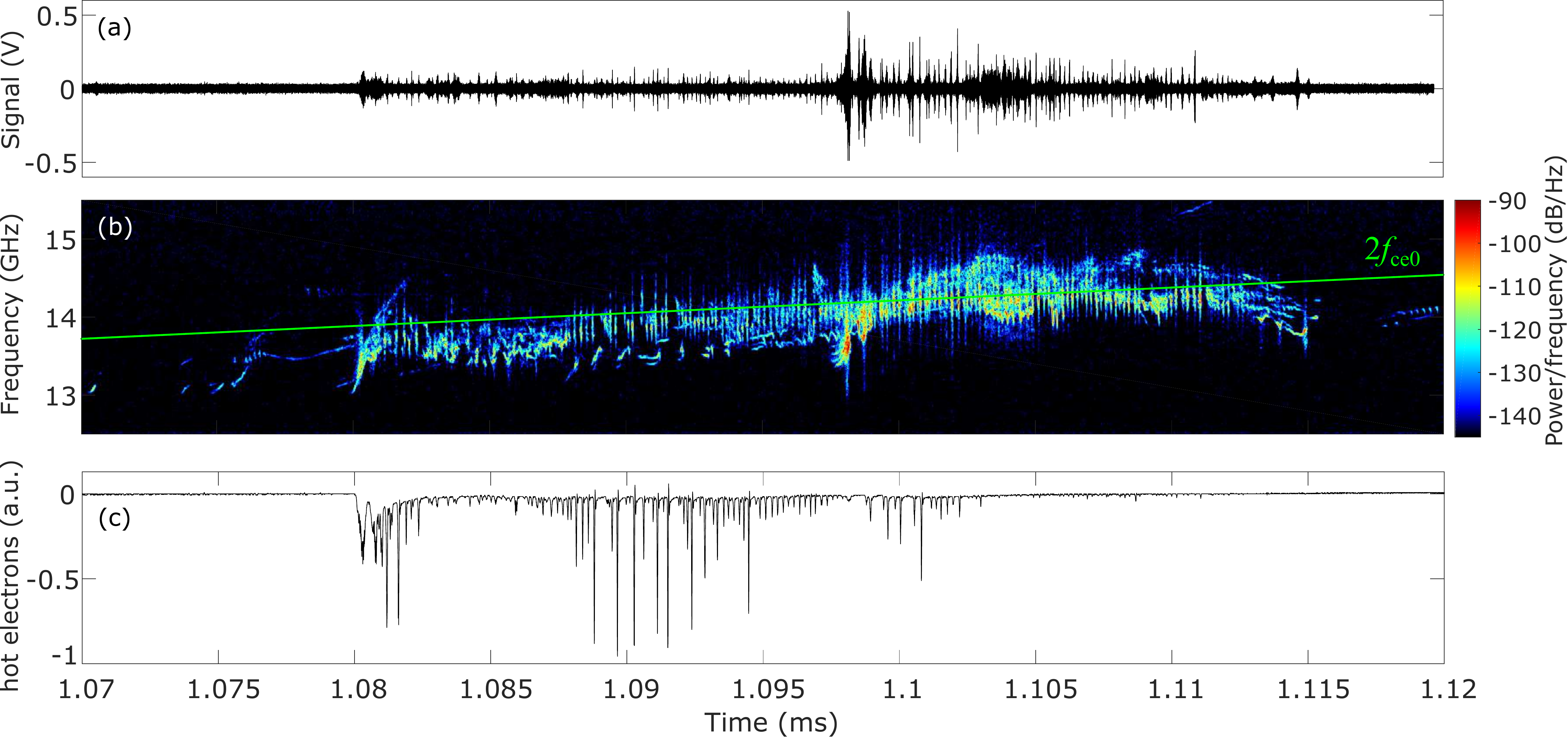}
	\caption{(a,b) Series of quasiperiodic bursts of plasma emission under double plasma resonance condition and (c) synchronous precipitations of energetic electrons from the trap. The value of $2f_{\mathrm{ce0}}$ at the trap center is shown by a green solid curve on the dynamic spectra.}
	\label{fig_spectrum4}
\end{figure*}

Very intense bursts observed just after ECR heating switch-off at the very beginning of the plasma decay phase have been attributed to excitation of the upper-hybrid  turbulence  under the double plasma resonance (DPR) condition \cite{viktorov_rf_dpr}. In this case, the instability growth rate of quasi-electrostatic plasma waves is greatly increased  when  the UHR frequency $\omega_{\mathrm{uh}}=(\omega_{\mathrm{pe}}^2+\omega_{\mathrm{ce}}^2)^{1/2}$ is equal to one of the harmonics of the electron gyrofrequency $\omega_{\mathrm{ce}}$. Due to further conversion of quasi-electrostatic waves into electromagnetic waves, this leads to the appearance of narrow-band emission near the harmonics of the electron gyrofrequency that lasts a finite time interval when the DPR condition is met. 

In the experiment, we observed DPR only at the second cyclotron harmonic which was limited by the bandwidth of used oscilloscope, see  Fig.~\ref{fig_spectrum} label \textit{d} and, in more details, Fig.~\ref{fig_spectrum4}. The DPR  condition, $\omega_{\mathrm{uh}}=2\omega_{\mathrm{ce}}$ or equivalently $\omega_{\mathrm{pe}}=\sqrt3\omega_{\mathrm{ce}}$, was met for the decreasing UHR frequency  during the plasma decay. For the example shown in Fig.~\ref{fig_spectrum4},  the instability was detected with a delay of about 80\,$\mu$s after the ECR heating switching-off as a long series of periodic  pulses (up to hundreds of bursts) of bright emission at frequencies near $2f_{\mathrm{ce0}}$. Synchronously, strong precipitations of fast electrons from the trap  are triggered. 
In contrast to previous types, this kind of emission has no frequency sweep in a single burst, the spectral width of radiation is $\Delta f/f\approx 0.1$. The period of oscillations is about 200\,ns and a single burst duration is about 50\,ns.


The electromagnetic energy released during this particular phase is more than a half of the total energy emitted during the entire decay phase.   By this moment of time, the density of the decaying cold plasma does not drop essentially, so that $N_{\mathrm{c}}>N_{\mathrm{h}}$ and $\omega_{\mathrm{pe}}>\omega_{\mathrm{ce}}$, hence we still consider the dense plasma case. 
The propagation direction of this radiation can not be clearly identified, we see equally strong signal in both antenna ports. 

It should be noted that increased UHR activity under the DPR conditions is not a rare event in astrophysical plasmas. It is related to the phenomenon of zebra patterns observed in the solar corona (type IV bursts) \cite{zlotnik_SP_1975_V43,zlotnik_SP_1975_V44}, to the decametric radiation of Jupiter  and even to the radio emissions of pulsars \cite{Zheleznyakon_2016_UFN,kuznetsov_2013}.

\subsection*{(e) X-mode emission in rarefied decaying plasma}

\begin{figure*}[tp]
	\includegraphics[width=1.0 \textwidth]{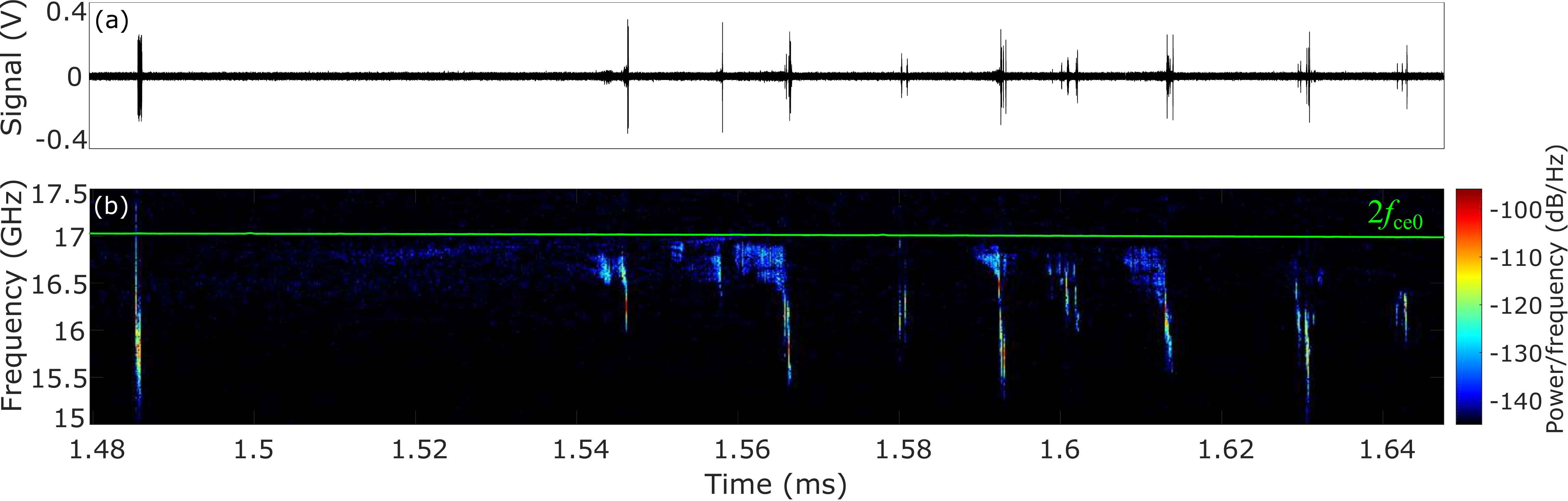}
	\caption{Series of quasiperiodic bursts of plasma emission in decaying plasma. The value of $2f_{\mathrm{ce0}}$ at the trap center is shown by a green solid curve on the dynamic spectra.}
	\label{fig_spectrum5}
\end{figure*}

Next type of instability develops at the mature phase of plasma decay when  $N_{\mathrm{h}}\sim N_{\mathrm{c}}$ and $\omega_{\mathrm{pe}}\ll\omega_{\mathrm{ce}}$, and the ECR heating of unstable electrons is off, see  Fig.~\ref{fig_spectrum} label \textit{e}  and Fig.~\ref{fig_spectrum5}. Such instability was first observed as quasi-periodic series of pulsed energetic electron precipitations that appeared with some delay after the heating power off \cite{Mansfeld_2007_JETP_maser}, later rough estimations of plasma cyclotron emission spectrum were obtained using microwave detectors \cite{viktorov_rf1}, and just recently detailed dynamic spectra were measured with a digital broadband oscilloscope \cite{viktorov_EPL}. 
In early experiments with detectors,  this radiation was observed presumably in the perpendicular to the magnetic field  direction; however this has not yet been confirmed in the latest experiments with fine-resolved spectra. 

A typical dynamic spectrum of plasma emission during the decay stage is shown in Fig.~\ref{fig_spectrum5}. Plasma emission consists of  a sequence of periodic bursts with decreasing frequency within a single burst. The duration of every burst is about 1\,$\mu$s and the period varies from 50 to 100\,$\mu$s. Almost every burst starts with a narrow-band emission at frequency slowly decreasing in time. Then during a very short period (about 100\,ns) the frequency of radiation rapidly drops with speed about 5\,GHz/$\mu$s.
The moment of fast frequency drop is correlated with energetic electrons precipitation from the trap ends. The frequency of emission is between $f_{\mathrm{ce0}}$ and $2f_{\mathrm{ce0}}$, and is also proportional to $f_{\mathrm{ce0}}$. 

Originally, the instability was interpreted as a result of resonant interaction at the fundamental cyclotron harmonic between the energetic electrons and the slow extraordinary waves propagating in a rarefied plasma across the external magnetic field at frequencies below the local gyrofrequency \cite{Mansfeld_2007_JETP_maser}. The delay between the ECR heating switch-off and triggering of the instability was explained by a polarization depression effect of the background (more dense and cold) plasma. However, more recent measurements clearly showed that the instability develops above the fundamental cyclotron harmonic near the trap center. It means that either the source region of this instability is not located  in the trap center, or the instability  can not be attributed to the slow X waves. The latter option is more reasonable since the region of the most uniform magnetic field in the trap center seems to be most favorable for excitation of kinetic EC instabilities. Thus, we assume now that this instability is more likely related to the  obliquely propagating fast X wave generated at the fundamental cyclotron harmonic, just as in the case of X-mode emission at the start-up phase, or its second harmonic. In most shots, we observe a delay before triggering of the instability at the fundamental harmonic that is caused by the  cut-off due to the presence of the cold and dense background plasma \cite{shalash_jetpl}. Interesting to note that, in spite of completely different mechanisms, the background plasma density required for triggering of the slow and fast X-modes at the fundamental harmonic is defined by qualitatively the same condition $\omega_{\mathrm{pe}}\lesssim\omega\sqrt{\varepsilon_{\mathrm{h}}/m_{\mathrm{e}}c^2}$  where  $\varepsilon_{\mathrm{h}}$ is the characteristic kinetic energy of radiating fast electrons. 

As discussed further, the sequences of pulsed bursts at the nonlinear instability phase may be explained in terms of a cyclotron maser model with a fast decrease of electromagnetic energy losses  \cite{shalash_2006}. The temporal behavior of microwave bursts and synchronous electron precipitations may be rather complex. Together with periodic patterns of bursts, single bursts may join in periodic groups of bursts, see Fig.~\ref{fig_spectrum5}, the interval between bursts may become irregular and generation may even be switched to a stochastic regime\cite{shalash_ppcf}. 

\subsection*{(f) Complex transients with  periodic frequency sweeps in decaying plasma }

\begin{figure}[tph]
	\includegraphics[width=85 mm]{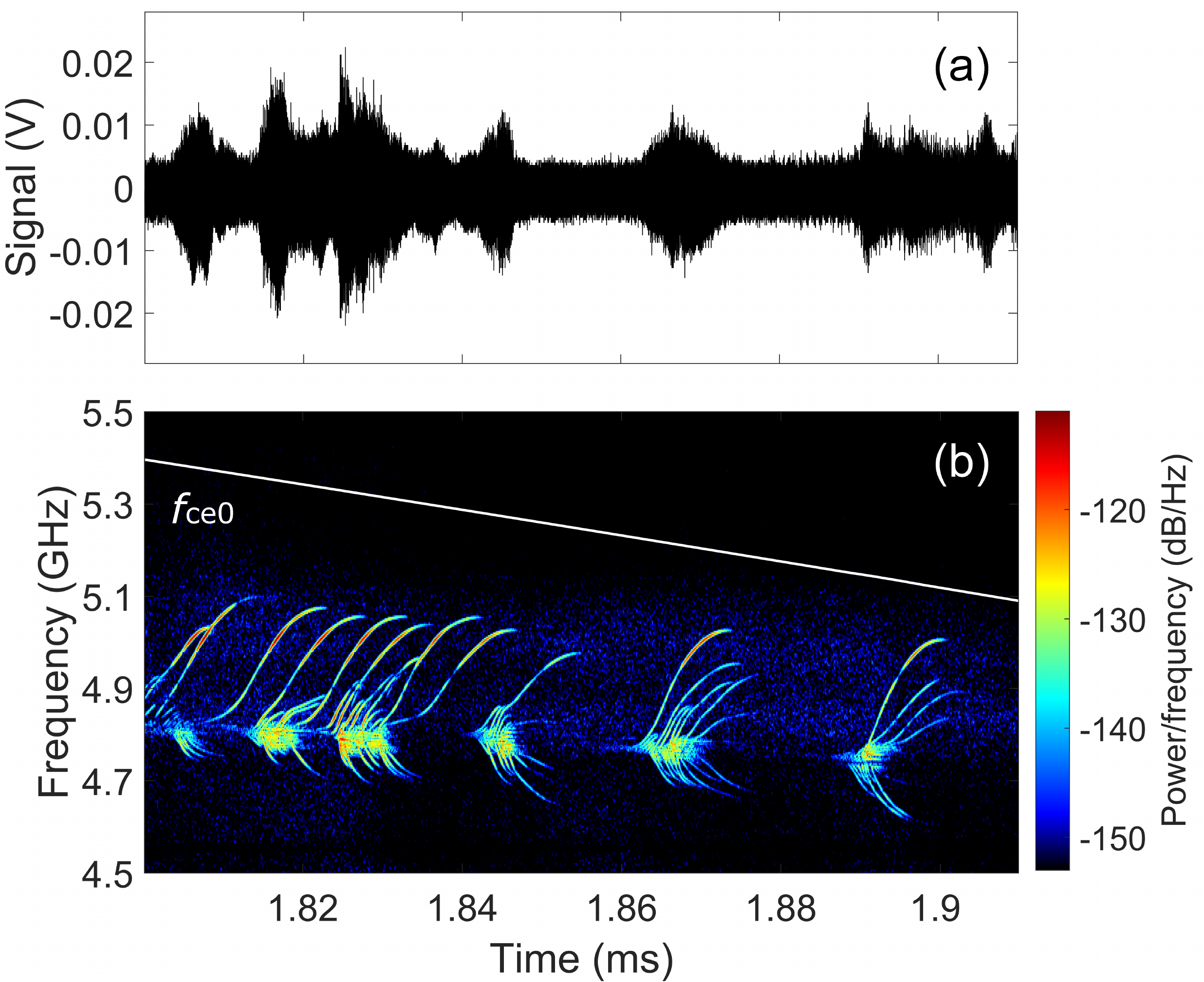}
	\caption{Electric field oscillations (a) and  corresponding dynamic spectrum (b) for the microwave emission during the late decay stage. }
	\label{fig_diplodoc}
\end{figure}

Highly transient complex spectra with many periodic frequency sweeps have been observed at  late phase of plasma  decay with $N_{\mathrm{h}} \gtrsim N_{\mathrm{c}}$ and  $\omega_{\mathrm{pe}}\ll\omega_{\mathrm{ce}}$, see Ref.~\onlinecite{Viktorov_2016_diplo}. 
A typical dynamic spectrum of the electric field oscillations in the excited wave is shown in Fig.~\ref{fig_diplodoc}. 
Excitation of these chirping wave packets is observed after ECR heating switch-off with a pronounced delay from 0.1 to 1\,ms and only when the ambient magnetic field is decreasing in time. Frequency band of the observed microwave emission is below the electron cyclotron frequency $f_{\mathrm{ce0}}$ in the trap center. The dependency of emission frequency band on the ambient magnetic field variations indicates that excited waves are of a cyclotron nature.
There is no reliable data on the propagation direction of this radiation.

The dynamic spectrum of emission is a set of highly chirped radiation bursts with both increasing and decreasing frequencies. Frequency variation rate is in the range 20--40\,MHz/$\mu$s. The bandwidth of this narrow-band emission is about $2\times 10^{-3} f_{\mathrm{ce0}}$. Rising tones experience in frequency sweep a saturation level after which the intensity of microwave emission is abruptly reduced. Wave packets are grouped in batches with the number of bursts in packs of about 10. The duration of narrow-band spike reaches about 10\,$\mu$s. Bundles of these pulses occur quasi-periodically with a period of 10--20\,$\mu$s. Sometimes a very long series of pulses with a duration up to 1\,ms are observed.

Unlike other types of instabilities observed in the experiment, in this case, electromagnetic emission in all cases is not accompanied by precipitating hot electrons. The energy of confined hot electrons may be estimated from its emission spectrum. Detuning between the emission frequency and the fundamental cyclotron frequency at the trap center is in the range 0.02--0.15 $f_{\mathrm{ce0}}$, what corresponds to the relativistic gyrofrequency shifts for electrons with characteristic energies 10--90 keV.

\section{Source of free energy for kinetic instabilities}

In  conditions of the reported experiment, the instability growth-rate in the kinetic regime may be calculated as an integral along the cyclotron resonance curve in a velocity space, see e.g. Ref.~\onlinecite{Wu85},
\begin{multline*}\gamma_{\mathrm{L}}=\int h(\upsilon_{||},\upsilon_\perp)\: \left(\pderv{f}{\upsilon_\perp}+\frac{k_{||}\upsilon_\perp}{\omega_{\mathrm{ce}}}\pderv{f}{\upsilon_{||}}\right) \times \\ \times\:\delta(\omega-k_{||}\upsilon_{||}-{n\omega_{\mathrm{ce}}}/\gamma)\:\mathrm{d}\upsilon_\perp\mathrm{d}\upsilon_{||}, 
\end{multline*}
where $f(\upsilon_{||},\upsilon_\perp)$ is the electron distribution function  over parallel and perpendicular velocities to the external magnetic field, $\delta$-function defines the relativistic resonance  condition, $k_{\parallel}$ is the wave-vector component along  the  magnetic field, $\gamma$ is the relativistic factor of electrons, $n$ is the cyclotron harmonic number, and ${h}$ is some mode-specific function. In case of the X mode propagating near the fundamental harmonic in rarefied plasma (most relevant to our study) $h=\pi^2\omega_{\mathrm{pe}}^2\omega_{\mathrm{ce}}\upsilon_{\perp}^2/4\omega$. 
We consider weakly relativistic electrons with energies less than $m_{\mathrm{e}}c^2$. In this case the derivative over parallel velocities may be neglected,   
$$\partial{f}/\partial{\upsilon_\perp}\gg({k_{||}\upsilon_\perp}/{\omega_{\mathrm{ce}}})\:\partial{f}/\partial{\upsilon_{||}}\sim(\upsilon_\perp/c)\:\partial{f}/\partial{\upsilon_{||}}\;.$$ 
Therefore, the instabilities are driven  presumably by positive slopes $\partial f/\partial \upsilon_\perp\!>\!0$ of the  distribution function, or, equivalently, by an inverse population of Landau levels. 
Below  we discuss  the formation of regions with $\partial f/\partial \upsilon_\perp\!>\!0$ in phase-space that may serve as a source of free energy for the kinetic instabilities. 

\begin{figure}[t]
\includegraphics[width=82 mm]{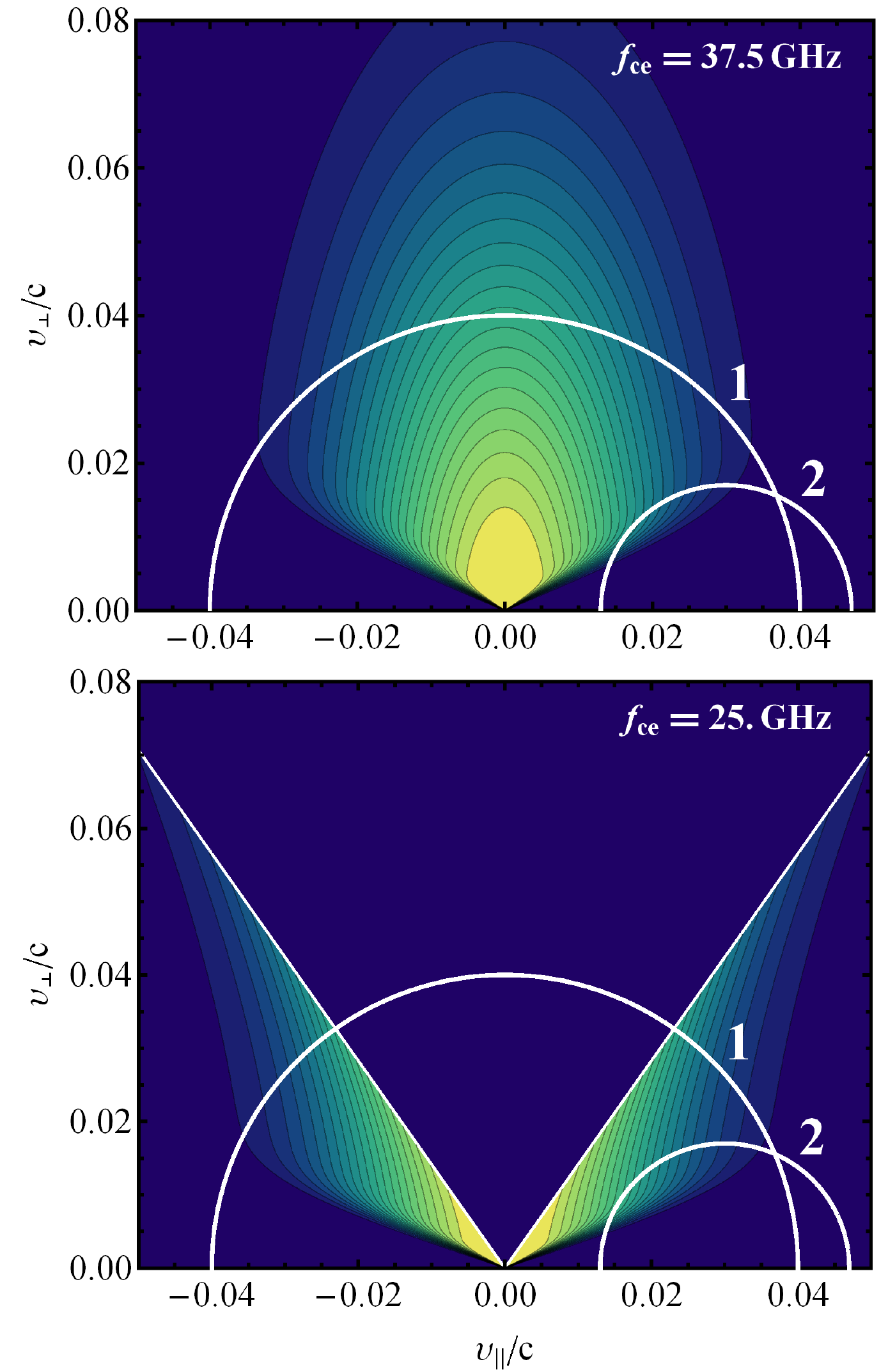}
\caption{Distribution function of fast electrons in velocity space in ECR heating region (top) and in a cross-section shifted in lower magnetic field towards the trap  center (bottom), and curves of the EC resonance corresponding to  transverse-propagating slow X wave at $\omega<\omega_{\mathrm{ce}}$ (1) and fast oblique X wave at $\omega>\omega_{\mathrm{ce}}$ (2). Note that for any propagation direction the relativistic resonance at the fundamental cyclotron harmonic always runs around the origin of coordinates for $\omega<\omega_{\mathrm{ce}}$ and outside the origin of coordinates for $\omega>\omega_{\mathrm{ce}}$.}
\label{fig_edf}
\end{figure}

As shown by many authors, strong ECR plasma heating in  adiabatic magnetic traps results in formation of anisotropic distributions of accelerated fast electrons characterized by a predominance of transverse velocities (relative to the direction of the magnetic field) to the longitudinal ones. For weakly relativistic electrons this may by modeled by bi-Maxwellian distribution, 
\begin {equation}\label {eq_edf} 
f=A\exp\left(-\frac{m_{\mathrm{e}} \upsilon_{||}^2}{2T_{||}}-\frac{m_{\mathrm{e}} \upsilon_{\perp}^2}{2T_{\perp}}\right)\Theta\left(\frac{\upsilon_{\perp}}{\upsilon}\right),
\end{equation}
where $\Theta(\upsilon_{\perp}/\upsilon)$ is an additional step-like function that describes the empty loss-cone in velocity space, $\upsilon_\perp/\upsilon<\sqrt{B/B_{\max}}$, typical for  adiabatically confined particles in a mirror trap. Isolines of this distribution function are shown in Fig.~\ref{fig_edf} (top plot).

One can see that regions with $\partial f/\partial \upsilon_\perp>0$ are only formed at the loss-cone boundary. This is favorable for excitation of the loss-cone instabilities in geometry where the resonance curve is entirely in the loss region in the velocity space and touches the loss-cone, see curve 2 in Fig.~\ref{fig_edf}. At the fundamental harmonic, such instabilities are only possible for inclined propagation at  $\omega>\omega_{\mathrm{ce}}$, i.e. for the fast X mode \cite{Melrose_1982}. Let us assume that the resonance curve touches the loss-cone at  velocities $\upsilon_{||}=\upsilon^\ast\cos\alpha^\ast$ and  $\upsilon_{\perp}=\upsilon^\ast\sin\alpha^\ast$.  This condition unambiguously defines the frequency and propagation angle for the most unstable X wave \cite{shalash_prl},
\begin {equation}\label {eq_lcc}
\omega/\omega_{\mathrm{ce}} =1+\upsilon_\ast^2/2c^2,
\quad k_{\vert \vert}c/\omega =\upsilon_\ast/(c\cos\alpha_\ast)
\end {equation}  
with  $\sin\alpha^\ast=\sqrt{B/B_{\max}}$  defined from the loss-cone boundary. This may be used to relate the frequency range of the measured emission and the energy of emitting electrons, see e.g. Sec.~III(a). 

Another plasma wave that may be  driven unstable by solely the anisotropy of velocity distribution  
is the whistler wave propagating at small angles to the magnetic field, see e.g. Ref.~\onlinecite{trakh_books}. This instability develops at frequencies $\omega<\omega_{\max}=(1-T_{||}/T_\perp)\;\omega_{\mathrm{ce}}$. On the other hand, in the experiments we measure $\omega_{\max}\sim\omega_{\mathrm{ce}}/2$   what corresponds to a very moderate level of the anisotropy required to excite the whistler instability.

A peculiar feature of our experiments is that ECR heating is shifted from the trap center towards the higher magnetic field. The distribution function of fast electrons in the ECR heating zone is approximated with Eq.~(\ref{eq_edf}); however, this distribution deforms while moving to lower magnetic field regions as shown in Fig.~\ref{fig_edf} (bottom plot). Such deformation may be understood assuming adiabatic motion of fast particles along the magnetic field lines such that the kinetic energy $\varepsilon=m_{\mathrm{e}}\upsilon^2/2$ and  the magnetic momentum $\mu=m_{\mathrm{e}}\upsilon_{\perp}^2/2B$ are conserved. 
Let the ECR heating corresponds to magnetic field $B=B_{\mathrm{ECR}}$ and distribution function $f(\upsilon_{||},\upsilon_\perp)$. Following Liouville's theorem, the distribution function at magnetic field $B'<B$ may be found as $f'(\upsilon'_{||},\upsilon'_\perp)=f(\upsilon_{||},\upsilon_\perp)$ where $\upsilon_{||}^2={\upsilon'_{||}}^2+{\upsilon'_{\perp}}^2(1-B/B')$ and $\upsilon_{\perp}^2={\upsilon'_{\perp}}^2 B/B'$. For the bi-Maxwellian  distributon (\ref{eq_edf}) one obtains
\begin {equation*}
f'=A\exp\left(-\frac{m_{\mathrm{e}} {\upsilon'_{||}}^{2}}{2T_{||}}-\frac{m_{\mathrm{e}} {\upsilon'_{\perp}}^2}{2T_{\perp}'}\right)\!\Theta'\left(\frac{\upsilon'_{\perp}}{\upsilon'}\right)\!\widetilde\Theta\left(\frac{\upsilon'_{\perp}}{\upsilon'}\right),
\end{equation*}
where  $\Theta'$ defines the empty loss-cone  $\upsilon'_\perp/\upsilon'<\sqrt{B'/B_{\max}}$ that corresponds to new magnetic field, $\widetilde\Theta$ defines the empty region  $\upsilon'_\perp/\upsilon>\sqrt{B'/B_{\mathrm{ECR}}}$ from which particles can not reach the ECR zone (the anti-loss-cone), and new effective transverse temperature is found from
$$\frac1{T'_{\perp}}=\frac1{T_{||}}\left(1-\frac {B_{\mathrm{ECR}}}{B'}\right)+\frac1{T_{\perp}}\frac {B_{\mathrm{ECR}}}{B'}.$$
Note that the effective temperature can be negative for small enough magnetic field: $T'_\perp<0$ when $B'<(1-T_{||}/T_\perp)\:B_{\mathrm{ECR}}$. So, a strong enough anisotropy of the fast electrons in the heating zone results in butterfly-like distribution in  central parts of the trap that possess $\partial f/\partial \upsilon_\perp>0$ everywhere except the boundary of the anti-loss-cone, see bottom panel in Fig.~\ref{fig_edf}. 

The described transition to ``globally unstable''   butterfly-like distribution function essentially effects the kinetic instabilities. First, it improves conditions for the fast X mode instability -- instead of touching the loss cone boundary at a singular point, the resonant curve may go through the body of the butterfly wing (compare curves 2 in the top and bottom panels of Fig.~\ref{fig_edf}). As a result, the fast X mode instability occurs in a much wider parameters range. Numerically we found that the maximum growth-rate occurs when the resonant curve touches the anti-loss-cone boundary. Thus we again can use Eqs.~(\ref{eq_lcc})  to relate the frequency and propagation angle of the measured emission to the energy of emitting electrons, however the angle $\alpha^\ast$ is now defined from the anti-loss-cone boundary as $\sin\alpha^\ast=\sqrt{B'/B_{\mathrm{ECR}}}$. 

Second, the butterfly  distribution seems to be less favorable for excitation of the slow X mode than the standard loss-cone distribution (\ref{eq_edf}). The reason is that for butterfly  distribution  there are always four sharp gradients along the resonant curve -- two of them correspond to the loss-cone  boundaries and are destabilizing ($\partial f/\partial \upsilon_\perp>0$), the other two  correspond to the anti-loss-cone  boundaries and are stabilizing ($\partial f/\partial \upsilon_\perp<0$), see curve 1 in Fig.~\ref{fig_edf} (bottom). The stabilizing gradients correspond to higher $\upsilon_\perp$, so they always dominate over destabilizing ones after the integration over the resonance curve. Thus, the slow X mode may develop only inside the ECR heating zone where the anti-loss-cone is absent. Since this region is highly localized,  the overall probability of the slow X mode instability is very low.

Assuming that $T_\perp\gg T_{||}$ and that spread in parallel velocities of fast electrons is defined by its Coulomb scattering on dense plasma during the ECR discharge, one obtains the following estimate:
$$T_{||}\approx\tau_{\mathrm{gyr}}\nu_{\mathrm{ei}}(T_\perp)\:T_\perp\approx12\;\mathrm{keV}/\sqrt{T_\perp[\mathrm{keV}]},$$
where $\tau_{\mathrm{gyr}}=1$\;ms is the gyrotron pulse length, and $\nu_{\mathrm{ei}}$ is the collision rate of fast electrons calculated for the background plasma density $N_e=10^{13}$\;cm$^{-3}$. In our conditions the mean electron energy $\varepsilon_{\mathrm{h}}\sim 10-100$ keV, thus the perpendicular temperature varies in the same range and the longitudinal temperature varies from $T_{||}\sim 1$\;keV for $T_{\perp}\sim 100$\;keV to $T_{||}\sim 4$\;keV for $T_{\perp}\sim 10$\;keV.  Correspondingly, the magnetic field variation needed for transition to  the globally unstable   butterfly-like distribution varies from  $\Delta B/B\sim1\%$ to 40\%, so in the worst case $B'=0.8$\;T. Having in mind that the minimum magnetic field in the trap center is always less than 0.8\,T, we can conclude that the globally unstable  distribution likely is formed  at all experimental conditions. 
Corresponded linear growth-rates of the X-mode instability are estimated in the range $\gamma_{\mathrm{L}}\sim  10^7-5\cdot 10^8$~s$^{-1}$ depending on both fast electron and background plasma densities.

\section{Cyclotron maser  paradigm}

As seen from the experimental data, the general picture of plasma emission due to kinetic instabilities is very complex. 
However, most of the observed features may be understood within a universal model of the cyclotron maser instability. It is based on a common description for the self-consistent evolution of particles and waves by means of the so-called quasilinear theory, a perturbative approach that involves many overlapped wave–particle resonances as a basis for diffusive particle transport in phase space \cite{QL00, QL001}. Quasilinear interaction of hot resonant electrons with the unstable electromagnetic wave reduces its transverse energy, at the same time the wave is exponentially increasing at the linear stage of the instability. As a result, some fast electrons fall within the loss cone and leave the trap. These losses reduce the instability growth rate and, finally, restrict the increase in the electromagnetic energy density in the system. 

The joint evolution of the electromagnetic field and resonant electrons may be described by the standard set of  quasilinear equations adopted for   magnetically confined inhomogeneous plasma\cite{trakh_books}:
\def\ee{|\mathcal{E}|_{\mathbf{k}}^2}
\def\jj{\mathbf{{J}}}
\def\djj{\mathrm{d}\jj}
\begin{equation}\label{eq:maser1}
\left\{\begin{aligned}
 &\pderv{F}{t} =\tilde{S}(\jj)+\mathcal{Q}, \quad\mathcal{Q}= \pderv{}{J_i} \left({\mathcal{D}_{ij}}\;\pderv{F}{J_j}\right),\\
        &\pderv{\ee}{t}-\mathbf{v}_{\mathrm{gr}}\pderv{\ee}{\mathbf{r}} = \left(\tilde\gamma_{\mathrm{L}}-\tilde\nu\right ) \ee\\
\end{aligned}\right.\end{equation}
with
\begin{gather*}
 \mathcal{D}_{ij}=D_{ij}(\jj)\int \ee\; \delta(\omega-k_{||}\upsilon_{||}-n\omega_{\mathrm{ce}}/\gamma)\;\mathrm{d} \mathbf{k},\\
 \tilde\gamma_{\mathrm{L}}=\int { {h_{i}(\jj)}\pderv{F}{J_i}\delta(\omega-k_{||}\upsilon_{||}-n\omega_{\mathrm{ce}}/\gamma)\;\djj}. 
\end{gather*}
These equations describe the wave intensity  $\ee$ and the hot electron distribution function $F(\jj)$ over the invariants of adiabatic motion $\jj=(\varepsilon,\;\mu)$, where $\varepsilon$ is the kinetic energy  and $\mu$ is the magnetic momentum. Here $\mathcal{Q}$ is the quasi-linear  operator responsible for the diffusion induced by unstable waves; $\tilde{S}$ is the source of hot particles, which itself may be modeled as a quasi-linear diffusion induced by an external (gyrotron) wave field; $\tilde\nu$ is the wave damping rate which is in most cases defined by Coulomb collisions in the background cold plasma; and $\tilde\gamma_{\mathrm{L}}$ is the instability growth rate  proportional to the gradient of the electron distribution function, so flattening of the distribution function due to the quasi-linear diffusion  results in saturation of the instability. 
Kernels of the diffusion coefficient, $D_{ij}$, and of the linear growth-rate,  $h_i$,  depend on a particular mode structure. 


Quasilinear equations are rather complex since they are formulated in at least three-dimensional space -- two invariants of the motion and one spatial coordinate in the direction of wave energy travel. Fortunately, the most essential physics may be described qualitatively using much simpler approach of balance equations: 
\begin{equation}\label{eq:bal}
\qquad\derv{N}{t} = S -\kappa EN\,,\quad  \derv{E}{t} = \left( \gamma_{\mathrm{L}}-\nu \right) E 
\end{equation}
with $$ \kappa=h/\varepsilon_{\mathrm{h}}, \quad\gamma_{\mathrm{L}}= hN, \quad  \nu=\tilde\nu+\tau_{\mathrm{gr}}^{-1}\ln R.$$
Here $N$ is the population of resonant electrons,  $E$ is the total energy of the unstable mode, $\kappa$ determines the losses of resonant electrons, $h$ is the growth-rate averaged over the wave propagation path, and $\nu$ is the total electromagnetic losses including not ideal reflection from the resonator ``ends'', where $\tau_{\mathrm{gr}}\sim L/v_{\mathrm{gr}}$ is the ``group time'' of wave packet propagation and $R$ is the reflection coefficient. The first equation describes the rf-field-induced losses of hot electrons with effective energy $\varepsilon_{\mathrm{h}}$, while the second equation describes the change of the wave energy in terms of the relation between the linear instability growth-rate $\gamma_{\mathrm{L}}$ and dissipation $\nu$. Although these equations looks very phenomenological, they may be formally obtained from a full set of quasilinear equations. To do that, one must assume a narrow and constant wave spectrum, $\ee\propto E(t)\,\delta({\mathbf{k}-\mathbf{k}_0})$,  seek the distribution function as an expansion over the eigenfunctions $f_n$ of the quasi-linear operator, 
$$f=\sum c_n(t)\, f_n(\jj),\quad\mathcal{Q}(f_n)=-\lambda_n f_n,$$ 
and consider only the slowest   $n=0$ term corresponding to the smallest $\lambda_n$. Then, the quasilinear equations equations (\ref{eq:maser1}) would result exactly in Eqs.~(\ref{eq:bal}) for $N=c_0(t)$. To calculate the coefficients $\kappa$ and $h$ it is necessary to know the solution of the original quasilinear kinetic equation. However, in our case it is possible to obtain a simple relation $\kappa = h/\varepsilon_{\mathrm{h}}$ by  imposing the energy conservation law in the from 
$$\derv{}{t}(\varepsilon_{\mathrm{h}} N+E) = \varepsilon_{\mathrm{h}} S -\nu E.$$  
Further details and comprehensive discussion on applicability of the balance equations can be found in Ref.~\onlinecite{trakh_books}.

Assume for simplicity that all the coefficients in Eq.~(\ref{eq:bal}) are either constant in time or adiabatically slowly vary. All oscillations in system (\ref{eq:bal}) decay with time and the system relaxes to the equilibrium state
$$
\overline{N}=\nu/h,\quad\overline{E}=S \varepsilon_{\mathrm{h}}/\nu.
$$
Linearizing the balance equations in the vicinity of this equilibrium  for the perturbations $\delta N,\delta E\propto \exp(\rmi\omega t)$ allowing for the assumption $\mathrm{Re}\,\omega\gg \mathrm{Im}\,\omega$, one  finds the complex frequency corresponding to the solution of the dispersion relation
$$\omega\approx\sqrt{Sh}+\rmi Sh/(2\nu).$$
In fact, the quantity  $1/\mathrm{Re}\,\omega$  determines the period of a  singular pulse  and $\mathrm{Re}\,\omega/\mathrm{Im}\,\omega$ determines the number of pulses in a series, i.e. both  parameters are experimentally obtained. Thus, comparing with the experimental data, we can estimate the value of the cyclotron instability growth rate at the linear instability threshold, $\gamma_{\mathrm{L}}=\nu$,  and the stationary density of the electromagnetic field energy referred to the hot electron energy as
$$
\gamma_{\mathrm{L}}\approx (\mathrm{Re}\,\omega)^2/(2\mathrm{Im}\,\omega),\quad \overline{E}/\varepsilon_{\mathrm{h}}\overline{N}\approx (2\mathrm{Im}\,\omega/\mathrm{Re}\,\omega)^2.
$$
The experimental data correspond to the growth rate $\gamma_{\mathrm{L}}\sim 10^8$\,s$^{-1}$ at the ECR start-up phase and for the under the DPR condition, and $\gamma_{\mathrm{L}}\sim 10^7$\,s$^{-1}$ for other stages. The maximum  electromagnetic energy density is about 1 $\mu$J/cm$^3$ for $\overline{N}\sim10^{10}$\,cm$^{-3}$ and  $\varepsilon_{\mathrm{h}}\sim300$ keV. In this case, the particle source amounts to $S\sim 10^{-3}\gamma_{\mathrm{L}} \overline{N}$.

\begin{figure*}[p]
{\small (a) Start-up phase: $S/\gamma_0 N_0=10^{-3}\left(1-\exp(-\tau/15000)\right),\;\; h/h_0=1,\;\;\nu/\gamma_0=1$}

\includegraphics[width=0.9 \textwidth]{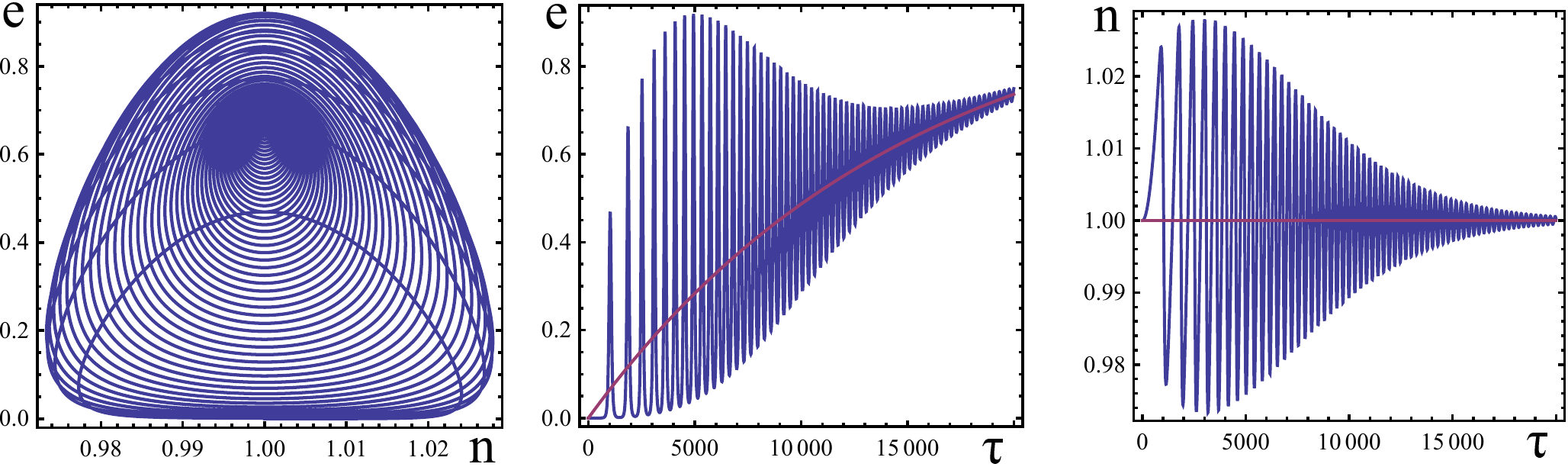}
\vspace{0.5pc}

{\small (b, c) Stationary ECR discharge: $S/\gamma_0 N_0=5\times 10^{-5},\;\; h/h_0=1,\;\;\nu=1$}

\includegraphics[width=0.9 \textwidth]{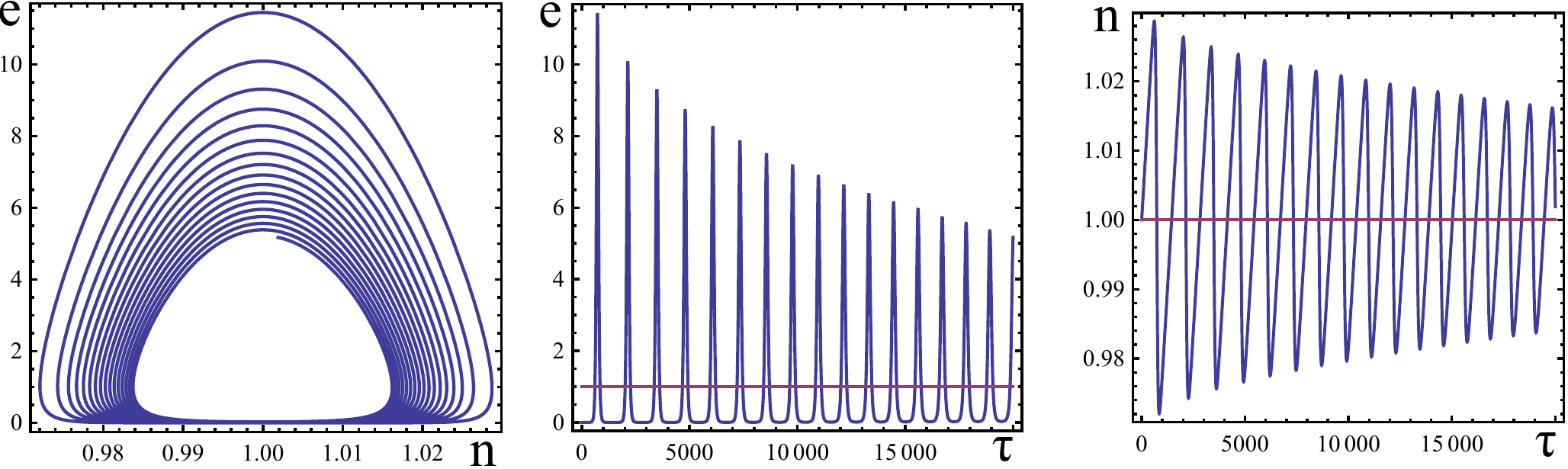}
\vspace{0.5pc}

{\small (d) Double-resonance condition: $S/\gamma_0 N_0=0,\;\; h/h_0=10\exp\left(-(\tau-10000)^2/1000^2\right),\;\;\nu/\gamma_0=1$}

\includegraphics[width=0.9 \textwidth]{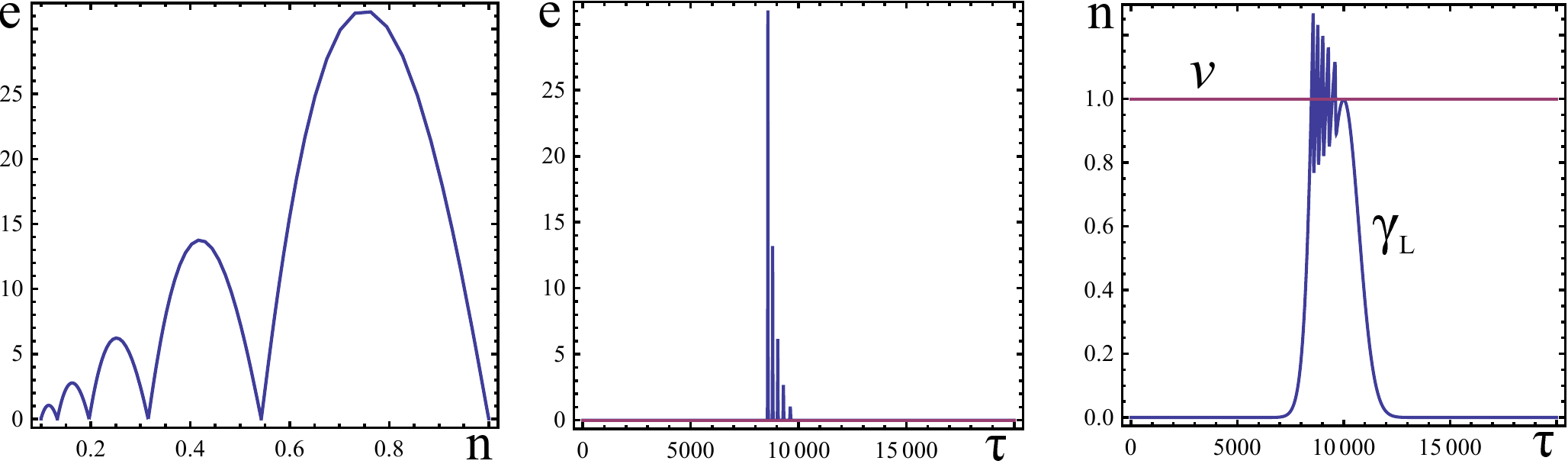}
\vspace{0.5pc}

{\small (e, f) Plasma decay: $S/\gamma_0 N_0=0,\;\; h/h_0=1,\;\;\nu/\gamma_0=1.6\exp\left(-\tau/10000\right)$}

\includegraphics[width=0.9 \textwidth]{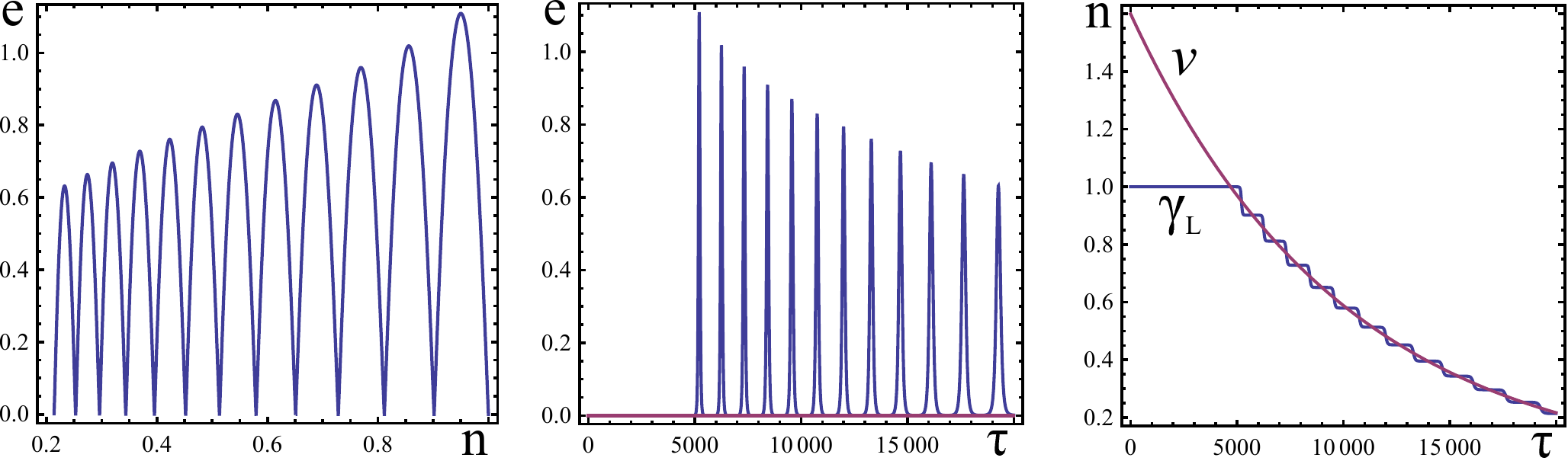}

\caption{Development of EC instability  in different  regimes simulated with the balance equations -- phase space $n-e$ (left panels), electromagnetic energy $e(\tau)$ and particle source $S(\tau)/S(0)$ (center), linear growth rate $\gamma_{\mathrm{L}}(\tau)/\gamma_0$ and dissipative losses $\nu(\tau)/\gamma_0$ (right). For calculations, the initial system (\ref{eq:bal}) was rewritten in dimensionless form allowing simulation in the absence of experimental data calibration when the absolute values of the measured fields and the particle flows are unknown: 
$    \dervi{n}{\tau} = \left(S/\gamma_0 N_0\right)-\left(h/h_0\right){e}n$,
$    \dervi{{e}}{\tau} = \left[\left(h/h_0\right)n-\left(\nu/\gamma_0\right)\right]{e}$.
Here, $N_0\approx \overline{N}$ is a certain characteristic density of the hot electrons, which, as opposed to  $\overline{N}$, does not vary in time,  $\gamma_0=h_0 N_0$ is the characteristic instability growth rate, $\tau=\gamma_0 t$, $n=N/N_0$, and ${e}=E/\varepsilon_{\mathrm{h}} N_0$.}
\label{fig:th}
\end{figure*}

For  more realistic simulations, we allow for the time variation of the particle source intensity $S$, the gain parameter $h$ and the radiation losses $\nu$. One can find that very simple balance equations have a rich family of oscillatory solutions, that depending on particular parameters may vary from harmonic oscillations in the vicinity of the equilibrium point  up to deeply modulated nonlinear waves with a big intermittency factor. Different regimes of oscillations have been successfully used to interpret the time-resolved burst activity in plasma emission and electron precipitations at different stages of discharge. Typical  examples of simulation of the instability dynamics  are shown in Fig.~\ref{fig:th}. 

\subsection*{(a) Start-up phase }
At the initial discharge stage, the free energy is gathered from the source of resonant particles $S(t)$ that is increasing in time after the ECR heating switch-on. Correspondingly, the equilibrium is also changing in time. This may explain the transition between two regimes of the fast X mode instability when periodic wave packets with fast raising frequency smoothly transforms into the continuous spectrum described in Sec.~III(a). The gradual increase of the intensity of  the fast electron source results in variation in the stationary level of radiation and related electron precipitations during the  discharge evolution. An increase of this level in times that significantly exceed the relaxation-oscillation period explains the ``quiet'' region (in which the burst activity is suppressed), experimentally observed between the solitary-peak oscillation and the stationary oscillation regions, see times $500-510$ $\mu$s in Fig.~\ref{fig_spectrum1}.

\subsection*{(b, c) Stationary ECR discharge}
During the stationary ECR discharge,  constant source $S$  maintains the equilibrium with a high enough wave energy, so the observed periodic and  quasi-stationary emissions may be attributed to small oscillations around this equilibrium. 
This picture fits well to the whistler wave instability observed below the central gyrofrequency $f_{\mathrm{ce0}}$ described in Sec.~III(b). However, microwave emissions observed in dense plasma at frequencies between $f_{\mathrm{ce0}}$ and $2f_{\mathrm{ce0}}$ as described in Sec.~III(c), show rather complex temporal and spectral evolution not suited to the proposed simple model. Theoretical interpretation of this phenomena is still a subject of our efforts.

\subsection*{(d) Double plasma resonance condition}
Just after the ECR heating switch-off, the source is absent $S=0$. However initial pool of resonant electrons is still full until the system is below the threshold, $\gamma_{\mathrm{L}}<\nu$. In Sec.~III(d) we assume the threshold is overcame at the DPR condition $\omega_{\mathrm{uh}}=2\omega_{\mathrm{ce}}$. In our model meeting of the DPR condition at some moment of time is described by a rapid increase of  gain $h$ in the balance equations (\ref{eq:bal}), that triggers the oscillatory motion, most likely, in a deeply non-linear regime of a giant impulse. 

In a more accurate description developed in Ref.~\onlinecite{viktorov_rf_dpr}, the pulsing mode of the UHR turbulence is related either to the competition of the instability and induced scattering of plasma waves, or the excitation of fast magnetosonic waves in the magnetic trap. The first interpretation seems more favorable due to a rather high growth rates of plasma waves typical of our experimental conditions. With Eqs.~(\ref{eq:bal}),  the induced scattering of the UH waves is described as additional non-linear modulation of the loss term $\nu$.

\subsection*{(e, f) Plasma decay}
At the late stage of  plasma decay, the source is also absent $S=0$, and the initial pool of resonant electrons may be empty because of  instabilities developed previously. The effective source of free energy is provided by a fast decrease of the wave damping rate, which constantly drives the system above the instability threshold. It has been shown that, even in the absence of a continuously acting source of nonequilibrium particles, the instability condition $\gamma_{\mathrm{L}}>\nu$ is recovered after each burst due to a monotonic decrease of the collisional absorption $\nu(t)\approx\nu_{\mathrm{ei}}$ in  a decaying plasma  \cite{shalash_2006}. In this way, we explain the sequences of pulsed bursts discussed in Sec.~III(e). 

Our simple model does not reproduce many features of a complex temporal behavior of microwave bursts and synchronous electron precipitations observed in the experiment. As already mentioned, together with periodic patterns of bursts, single bursts may join in periodic groups of bursts with   irregular intervals between the patches of bursts. 
To describe more complex  maser dynamics, the balance equations may be further generalized to include several competing unstable modes. In particular, some features in the observed complex temporal dynamics in a decaying plasma, such as burst grouping and spontaneous transition from  quasi-periodic to stochastic regimes,  was explained with the idea of self-modulation of a maser due to interference of two counter-propagating degenerate unstable waves resulting in spatial modulation of amplification \cite{shalash_ppcf}. However, this model can not describe the frequency sweeps resolved in the reported experiments, and especially the complex transients in dynamic spectra described in Sec.~III(f). 

Modification of the spectra is typically  related to a neat modification of the distribution function of the resonant particles.  Within quasi-linear approach, which is most developed for the whistler and Alfven wave cyclotron instabilities in space \cite{trakh_books}, the frequency sweeping may be explained as a result of relatively slow modification of the average distribution function of the resonant particles in the frame of the full kinetic equation. Although the quasilinear theory is in principle able to describe fast events, such as switching from kinetic to hydrodynamic instability in the inner magnetosphere of the Earth \cite{trakh2,trakh3}, it faces essential difficulties in many cases involving fast transients.  

\section{Holes and clumps paradigm}

To explain complex transients with periodic frequency sweeps observed in decaying plasma as described in Sec.~III(f), we consider 
an alternative universal physical mechanism based on formation of nonlinear phase space structures in the proximity of the wave-particle resonances of a kinetically unstable bulk plasma mode\cite{Viktorov_2016_diplo}. Such regimes were proposed as a possible mechanism of generation of narrow-band chorus emissions in the Earth's magnetosphere\cite{trakh4,trakh5}. Later evolution of similar phase space structures was studied in Refs.~\onlinecite{berk96,berk97, brei97} for the case of essential wave dissipation, which is more relevant to our experiments. 
Here the wave resonances do not overlap, and the global quasi-linear transport in phase space is suppressed. As the mode grows, most of the particles respond adiabatically to the wave, and only a small  group of resonant particles mix and cause local flattening of the distribution function in phase space within or near the separatrices formed by the waves. However, when a linear dissipation from a background plasma is present, the saturated plateau state becomes unstable, and the mode tends to grow explosively. This results in  the formation and subsequent evolution of long-living structures in the particle distribution, so-called holes (a depletion of particles) and clumps (an excess of particles), 
whose frequencies are slightly up- and down-shifted with respect to that of the initial instability. Their subsequent convective motion in the phase space is synchronized to the change in wave frequency, thus leading to complex chirping patterns in dynamical spectra of unstable waves\cite{lil10,nyq12,nyq13}. 

This model seems to be the most suitable to explain the transients with  periodic frequency sweeps observed in our experiment. For a very rough estimate, one can isolate one spectral component and  model the  evolution in time of one chirping event. 
Then, in the collisionless limit for resonant particles, a bounce average of  Maxwell--Vlasov kinetic equations yields the frequency shift $\delta\omega$ an unstable mode as\cite{berk97}
\begin{equation}
\label{eq:bb}
\qquad \delta\omega\approx \frac{16\sqrt2}{3\sqrt{3}\pi^2}\,\gamma_{\mathrm{L}}\sqrt{\nu t}\,.
\end{equation}
This expression is valid for the regime where resonant particles are deeply trapped in the wave  potential and perturbation of the passing particle is negligible\cite{berk97}. Although this result was originally obtained for purely electrostatic and one-dimensional bump-on-tail instability, it gives reasonable qualitative estimates for more complex electromagnetic problems\cite{pinches04,les10}. 
Thus, we may apply Eq.~(\ref{eq:bb}) to our case of the  extraordinary wave propagating in rarefied plasma. 
Fitting the the measured frequency sweeps $\delta\omega=\sqrt{A\:t}$ by constant  $A$ and assuming near-threshold condition $\gamma_{\mathrm{L}}\approx\nu$, one may estimate the upper boundary for the dissipation $\nu$ or, equivalently, the lower boundary for the  growth rate $\gamma_{\mathrm{L}}$ compatible with Eq.~(\ref{eq:bb}). In the experiments we obtain typical values  $A=(1-3)\times10^{21}$\,s$^{-3}$ resulting in 
$\gamma_{\mathrm{L}}\approx(1.7-2.5)\times10^7$\,s$^{-1}$, what is in reasonable agreement with independent estimates.  More details may be found in Ref.~\onlinecite{Viktorov_2016_diplo}.

\section{Summary}

Studies of cyclotron instabilities in magnetically trapped laboratory plasmas have a long history but remain topical, mostly with the advent of powerful sources of microwave radiation (especially, gyrotrons), which allow to sufficiently raise the level of energy input into the plasma thereby increasing the ``energetics'' of nonequilibrium resonant particles. 
An important advantage of nonequilibrium ECR discharge plasma  is an opportunity to recreate different conditions for excitation and amplification of waves in plasma in the same setup. On the other hand, generation of pulsed electromagnetic radiation accompanied by hot electrons precipitation in the laboratory conditions has much in common with similar processes occurring in the magnetosphere of Earth, planets, and solar coronal loops. Thus, this paper may be of interest in the context of a laboratory modeling of non-stationary processes of wave-particle interactions in space plasma, since there are a lot of open questions about the origin of some types of emissions in space cyclotron masers, especially mechanisms of fine spectral structures. 


\begin{acknowledgments} The work has been supported by RFBR (grants No. 15--32--20770, 16--32--60056, 16--02--00625).
The authors thank Tektronix Inc.\;and Keysight Technologies Inc. for their technical support.
\end{acknowledgments}

\section*{References}

\end{document}